\documentclass[11pt]{article}

\begin{document}

\newcommand{\half}{\frac{1}{2}}
\newcommand{\inv}{^{-1}}

\title{Drinfeld-Manin Instanton and Its Noncommutative Generalization}
\author{Yu Tian \\{\it Institute of Theoretical Physics, Chinese Academy of Sciences}\\
{\it P. O. Box 2735, Beijing 100080, China}\\
{\tt ytian@itp.ac.cn}}

\maketitle

\begin{abstract}
The Drinfeld-Manin construction of $U(N)$ instanton is
reformulated in the ADHM formulism, which gives explicit general
solutions of the ADHM constraints for $U(N)$ ($N\ge 2k-1$)
$k$-instantons. For the $N<2k-1$ case, implicit results are given
systematically as further constraints, which can be used to the
collective coordinate integral. We find that this formulism can be
easily generalized to the noncommutative case, where the explicit
solutions are as well obtained.
\end{abstract}

\newpage

\section{Introduction}

Instanton solutions in gauge field theory are of great physical
and mathematical interest \cite{Coleman, Rajaraman, S S, S V, F U,
D K}. Many significant achievements have been made in this region
since their discovery in 1975 \cite{BPST}.

Because of the great significance of instanton solutions in
various aspects of physics and mathematics, it is necessary to
obtain all these solutions in gauge field theory. This task was
almost accomplished in 1978, when Atiyah, Hitchin, Drinfeld and
Manin (ADHM) established the famous construction of instantons for
almost all gauge groups\footnote{More precisely, the construction
for exceptional groups is not known yet.} \cite{ADHM, C W S}. This
ADHM construction essentially reduces the problem of solving a set
of nonlinear partial differential equations, which defines the
instantons, to that of solving a set of quadratic algebraic
equations, called the ADHM constraints. It gives the most general
instanton configurations, and so provides the probability to learn
the whole instanton moduli spaces.

But even algebraic equations are not always solvable, so the ADHM
constraints remain a difficult problem. In other words, it is hard
to attain satisfactory parametrization of instanton moduli spaces.
For gauge group $U(N)$, or essentially $SU(N)$, during a rather
long time since the presentation of ADHM construction, general
solutions of the ADHM constraints are known only when $k=1$ and
$N$ arbitrary or $k\leq 3$ and $N=2$ \cite{C W S, K Sh, Tian4}
(except for the Drinfeld-Manin parametrization explained below),
where $k$ is the topological charge, or equivalently the instanton
number \cite{Tian3}, which is an integer classifying the instanton
solutions. In 1999, Dorey \textit{et al.} essentially rediscovered
the Drinfeld-Manin parametrization for $N\ge 2k$ \cite{DHKMV}, of
which they seemed not aware.

In recent years the study of gauge field theory on noncommutative
space time becomes an active research area \cite{K S, D N, Szabo},
mostly due to its relevance with string theory \cite{S W}. An
interesting phenomenon in noncommutative gauge field theory is
that instanton solutions survive the space-time noncommutativity,
and the moduli spaces of them get even better behaved
\cite{Nakajima}. Correspondingly, the ADHM construction has been
generalized to the noncommutative case \cite{N S,
Tian2}.\footnote{In fact, the ADHM constraints arise naturally as
the D-flat condition of the worldvolume theory of the D$p$-brane
in D$p$-brane-D$(p+4)$-brane bound systems \cite{Witten, Douglas}.
When a constant NS-NS $B$-field is present in the worldvolume of
the D$(p+4)$-branes, the worldvolume theory of the D$(p+4)$-branes
becomes noncommutative, and a Fayet-Iliopoulos D-term appears in
the worldvolume theory of the D$p$-branes \cite{Moore}.
Corresponding to this term, one must add a constant term to the
ADHM constraints.} The noncommutative ADHM constraints seem even
more difficult to solve: for gauge group $U(N)$, up to now only
when $k=1$ and $N$ arbitrary or $k=2$ and $N=1$ general solutions
are known \cite{LTY}.

Drinfeld and Manin presented another construction of instantons
\cite{D M} shortly after the ADHM construction, from a slightly
different point of view. This construction explicitly gives
parametrization of the $U(2k)$ $k$-instanton moduli space. In
addition, all $U(N)$ $k$-instanton configurations can be
indirectly obtained. Their original description of this
construction was in a vector-bundle language. In this article we
will translate it into the more familiar ADHM language and see how
they give explicit general solutions of the ADHM constraints with
gauge group $U(N)$ ($N\ge 2k-1$) and topological charge $k$. For
the $N<2k-1$ case, the further constraints are hard to solve
explicitly, but our systematic discussion can offer an indirect
way to the collective coordinate integral in this case. More over,
fortunately, a noncommutative generalization of this ADHM
formulation of Drinfeld-Manin instanton is straightforward.

This paper is organized as follow. In Sec.\ref{ord} and
Sec.\ref{NC} we recall the definition of instantons and the ADHM
construction, in the commutative case and the noncommutative case,
respectively. In Sec.\ref{DM} the Drinfeld-Manin construction is
briefly reviewed and reformulated in the ADHM formulism. This
construction is generalized to the noncommutative case in
Sec.\ref{NCDM}. In the appendix, the conditions for a Hermitian
matrix of restricted rank are given. These conditions are needed
in the discussion of the $N<2k$ case.

\section{Instantons and (ordinary) ADHM construction}\label{ord}

Instanton solutions in (Euclidean) gauge field theory were
discovered by Belavin, Polyakov, Schwartz and Tyupkin (BPST) in
1975 \cite{BPST}. They are defined by the so-called
(anti-)self-dual equations:
\begin{equation}\label{instanton}
\tilde{F}_{mn}=\pm F_{mn},\quad(m,n=1,2,3,4)
\end{equation}
and the solutions are known as self-dual (SD, for ``+'' sign) and
anti-self-dual (ASD, for ``$-$'' sign) instantons. The definition
of dual field $\tilde{F}_{mn}$ is familiar in electrodynamics,
which is
\begin{equation}
\tilde{F}_{mn}=\half\epsilon_{mnpq}F_{pq}
\end{equation}
when the standard Euclidean metric $g_{mn}=\delta_{mn}$ is
assumed. We note that the notions of SD and ASD are interchanged
by a parity transformation. Without loss of generality we will
consider only the ASD instantons.

All the (ASD) instanton solutions can be obtained by the ADHM
construction \cite{ADHM, C W S}, as follows. In this construction
we introduce the following ingredients (for $U(N)$ gauge theory
with instanton number $k$):
\begin{itemize}
\item $k\times k$ matrix $B_{1,2}$, $k\times N$ matrix $I$ and
$N\times k$ matrix $J$,
\item the following quantities:
\begin{eqnarray}
\mu _{r} &=&[B_{1},B_{1}^{\dagger }]+[B_{2},B_{2}^{\dagger
}]+I\,I^{\dagger
}-J^{\dagger }J,  \label{ADHM1} \\
\mu _{c} &=&[B_{1},B_{2}]+I\,J.  \label{ADHM2}
\end{eqnarray}
\end{itemize}
The claim of ADHM is as follows:
\begin{itemize}
\item Given $B_{1,2}$, $I$ and $J$ such that $\mu_r=\mu_c=0$, an
ASD gauge field can be constructed;
\item All ASD gauge fields can be obtained in this way.
\end{itemize}

It is convenient to introduce a quaternionic notation for the
4-dimensional Euclidean space-time indices:
\begin{equation}
x\equiv x^{n}\sigma_{n},\quad\bar{x}\equiv
x^{n}\bar{\sigma}_{n}=x^\dag,
\end{equation}
where $\sigma_{n}\equiv(i\vec{\tau},1)$ and $\tau^{c}$, $c=1,2,3$
are the three Pauli matrices, and the conjugate matrices
$\bar{\sigma}_{n}\equiv(-i\vec{\tau},1)=\sigma_{n}^{\dag}$. Then
the basic object in the ADHM construction is the $(N+2k)\times 2k$
matrix $\Delta $ which is linear in the space-time coordinates:
\begin{equation}
\Delta=a+b\bar{x}\equiv a+b(\bar{x}\otimes 1_k),  \label{Delta}
\end{equation}
where the constant matrices
\begin{equation}\label{canonic}
a=\left(\begin{array}{cc}
I^{\dag } & J \\
B_{2}^{\dagger } & -B_{1} \\
B_{1}^{\dagger } & B_{2}
\end{array}\right),
\quad b=\left(\begin{array}{cc}
0_{N\times k} & 0_{N\times k} \\
1_k & 0 \\
0 & 1_k
\end{array}\right).
\end{equation}
It is easy to check that the ADHM constraints (\ref{ADHM1}) and
(\ref{ADHM2}) are equivalent to the so-called factorization
condition:
\begin{equation}  \label{factorize}
\Delta^\dagger\Delta=\left(\matrix{f^{-1} & 0 \cr 0 &
f^{-1}}\right),
\end{equation}
where $f(x)$ is a $k\times k$ Hermitian matrix. From the above
condition we can construct a Hermitian projection operator
$P$:\footnote{We use the following abbreviation for expressions
with $f$: $$\Delta f \Delta^\dag \equiv \Delta \left(\matrix{f & 0
\cr 0 & f}\right) \Delta^\dag =\Delta (1_2\otimes
f)\Delta^\dag.$$}
\begin{equation}
P=\Delta f\Delta^\dag.
\end{equation}

Obviously, the null space of $\Delta^\dag(x)$ is of $N$ dimension
for generic $x$. The basis vector for this null space can be
assembled into an $(N+2k)\times N$ matrix $U(x)$:
\begin{equation}
\Delta^\dag U=0,
\end{equation}
which can be chosen to satisfy the following orthonormal
condition:
\begin{equation}  \label{normal}
U^\dag U=1.
\end{equation}
The above orthonormal condition guarantees that $UU^\dag$ is also
a Hermitian projection operator. Now it can be proved (see
\cite{Tian2}) that the completeness relation
\begin{equation}  \label{complete}
P+UU^\dag=1
\end{equation}
holds if $U$ contains the whole null space of $\Delta^\dag$. In
other words, this completeness relation requires that $U$ consists
of all the zero modes of $\Delta^\dag$.

The (anti-Hermitian) gauge potential is constructed from $U$ by
the following formula:
\begin{equation}\label{A by U}
A_m= U^\dag\partial_m U.
\end{equation}
Then we get the corresponding field strength:
\begin{eqnarray}
F_{mn}&=&\partial_{[m} A_{n]} + A_{[m} A_{n]} \equiv
\partial_m A_n - \partial_n A_m + [A_m,A_n]  \nonumber \\
&=&\partial_{[m}(U^\dag\partial_{n]}U)
+(U^\dag\partial_{[m}U)(U^\dag\partial_{n]}U)
=\partial_{[m}U^\dag(1-UU^\dag)\partial_{n]}U  \nonumber \\
&=&\partial_{[m}U^\dag\Delta f\Delta^\dag\partial_{n]}U
=U^\dag\partial_{[m}\Delta f\partial_{n]}\Delta^\dag U =U^\dag
b\bar\sigma_{[m}\sigma_{n]}f b^\dag U  \nonumber \\
&=& 2i\bar\eta^c_{mn}U^\dag b(\tau^c f)b^\dag U.  \label{F by U}
\end{eqnarray}
Here $\bar\eta^c_{mn}$ is the standard 't Hooft $\eta$-symbol,
which is anti-self-dual:
\begin{equation}
\frac{1}{2}\epsilon_{mnpq}\bar\eta^c_{pq}=-\bar\eta^c_{mn}.
\end{equation}

\section{Noncommutative ADHM construction}\label{NC}

First let us recall briefly the gauge field theory on
noncommutative Euclidean space (time)\footnote{For general reviews
on noncommutative geometry and field theory, see, for example,
\cite{K S, D N, Szabo, Nekrasov}.}. For a general noncommutative
$\mathbf{R}^4$ we mean a space with Hermitian-operator coordinates
$x^{n}$, $n=1,\cdots,4$, which satisfy the following relations:
\begin{equation}
[x^{m},x^{n}]=i\theta^{mn},
\end{equation}
where $\theta^{mn}$ are real constants. If we assume the standard
(Euclidean) metric for the noncommutative $\mathbf{R}^4$, we can
use the orthogonal transformation with positive determinant to
change $\theta^{mn}$ into the following standard form:
\begin{equation}\label{theta}
(\theta^{mn})=\left(
\begin{array}{cccc}
0 & \theta^{12} & 0 & 0 \\
-\theta^{12} & 0 & 0 & 0 \\
0 & 0 & 0 & \theta^{34} \\
0 & 0 & -\theta^{34} & 0
\end{array}
\right).
\end{equation}
By using this form of $\theta^{mn}$, the only non-vanishing
commutators are
\begin{equation}
[x^{1},x^{2}]=i\theta^{12},\quad[x^{3},x^{4}]=i\theta^{34},
\end{equation}
and the other two obtained by using the anti-symmetric property of
commutators.

The full noncommutative gauge field theory demands most of the
abstract notions from noncommutative geometry, such as
differential forms and vector bundles on noncommutative spaces
\cite{Connes, Madore}. But for the $U(N)$ gauge theory on
noncommutative Euclidean space, things will be much simpler: in
fact, the final effect is almost to replace all the coordinates in
ordinary $U(N)$ gauge theory with the above operator coordinates.
However, a definition of derivatives in the noncommutative case
are necessary for any gauge field theory. We define
\begin{equation}
\partial_m f \equiv - i \theta_{mn} [x^n, f],
\end{equation}
where $\theta_{mn}$ is the matrix inverse of $\theta^{mn}$. For
our standard form (\ref{theta}) of $\theta^{mn}$ we have
\begin{equation}
\partial_1 f = {\frac{i }{\theta^{12}}} [x^2,f], \quad
\partial_2 f = - {\frac{i }{\theta^{12}}} [x^1,f],
\end{equation}
and similar relations for $\partial_{3,4}$.

Now we recall the noncommutative ADHM construction \cite{N S}
briefly here. By introducing the same data as above but
considering the coordinates as noncommutative we see that the
factorization condition (\ref{factorize}) still gives $\mu_c=0$,
but $\mu_r$ no longer vanishes. It is easy to check that the
following relation holds:
\begin{equation}
\mu_r=\zeta\equiv 2\theta^{12}+2\theta^{34}.
\end{equation}
The form (\ref{factorize}) of ADHM constraints is invariant
whether the space time is commutative or not.

The space-time noncommutativity brings nontrivial effects on the
physics of gauge field theory. A remarkable example is the mixing
between the infrared (IR) and the ultraviolet (UV) degrees of
freedom \cite{M R S}. Concerning the ADHM construction, in the
noncommutative case the operator $\Delta^\dag\Delta$ always has no
zero mode (see \cite{K S}) and the moduli spaces of noncommutative
instantons are better behaved than their commutative counterparts
(see, for example, the lectures by H. Nakajima \cite{Nakajima}). A
related interesting fact is that noncommutative $U(1)$ gauge
theory allows nonsingular instanton solutions \cite{N S,
Furuuchi}, while in the commutative case the simplest gauge group
for which nonsingular instanton solutions exist is $SU(2)$.

Whether in the commutative case or in the noncommutative case, we
can find that the above ADHM construction with $b$ in the canonic
form (\ref{canonic}) is unaffected by the following
transformations:
\begin{equation}\label{aux}
\Delta\rightarrow\left(\begin{array}{cc}
1_N & 0 \\
0 & 1_2\otimes u \\
\end{array}\right)\Delta(1_2\otimes u^\dag),
\end{equation}
where $u\in U(k)$. This is called the auxiliary symmetry of the
ADHM construction, which acts on $a$, $f$ and $U$ as
\begin{eqnarray}
B_1 &\rightarrow& u B_1 u^\dag, \\
B_2 &\rightarrow& u B_2 u^\dag, \\
I &\rightarrow& u I, \\
J &\rightarrow& J u^\dag, \\
f &\rightarrow& u f u^\dag, \\
U &\rightarrow& \left(\begin{array}{cc}
1_N & 0 \\
0 & 1_2\otimes u \\
\end{array}\right)U.
\end{eqnarray}
Now we can do a parameter counting for the (commutative or
noncommutative) ADHM $U(N)$ $k$-instanton. $a$ in the form
(\ref{canonic}) contains $4k^2+4Nk$ real parameters. The ADHM
constraints (\ref{ADHM1},\ref{ADHM2}) impose $3k^2$ real
conditions on them, and the auxiliary symmetry removes further
$k^2$ real degrees of freedom. In total we have $4Nk$ real
parameters left, which is expected according to physical analysis
\cite{C W S}.

The above ADHM construction is also unaffected by the following
transformations:
\begin{equation}\label{global}
\Delta\rightarrow\left(\begin{array}{cc}
\mathcal{U} & 0 \\
0 & 1_{2k} \\
\end{array}\right)\Delta,\quad\mathcal{U}\in SU(N),
\end{equation}
which can be regarded as the global gauge rotations of the
instanton configuration. This global gauge symmetry leaves
$B_{1,2}$ and $f$ unchanged and acts on $I$, $J$ and $U$ as
\begin{eqnarray}
I &\rightarrow& I\mathcal{U}^\dag \\
J &\rightarrow& \mathcal{U}J \\
U &\rightarrow& \left(\begin{array}{cc}
\mathcal{U} & 0 \\
0 & 1_{2k} \\
\end{array}\right)U.
\end{eqnarray}
If we wish to eliminate this global gauge symmetry from the $4Nk$
real parameters and retain the ``purely'' physical degrees of
freedom, the number of independent real parameters will be
$4Nk-N^2+1$ for $k\geq N/2$, and $4Nk-N^2+(N-2k)^2+1=4k^2+1$ for
$K\leq N/2$ because in this case only $N^2-(N-2k)^2-1$ degrees of
freedom in the $SU(N)$ group act nontrivially on $I$ and $J$.

\section{ADHM formulation of the Drinfeld-Manin construction}\label{DM}

Shortly after the ADHM construction was established, Drinfeld and
Manin successfully constructed all instanton solutions from a
so-called ``instanton bundle'' point of view \cite{D M}, which we
call the Drinfeld-Manin construction. In this construction, the
Euclidean space time is compactified by a point to $S^4$ and the
instanton gauge potentials are considered as Levi-Civita
connections on some nontrivial vector bundles, named instanton
bundles, on this $S^4$. The instanton bundles are complex bundles
(for the case of $U(N)$ gauge group) orthogonally complementary,
under some metrics, to a trivial vector bundle $M$. The
(anti-)self-duality of the Levi-Civita field strength imposes some
conditions on the metric, which are actually the ADHM constraints.

We can always perform a complex linear transformation (on the
basis vectors of the fibre space) to make the (Hermitian) metric
standard. If we have done so, then the column vectors of $\Delta$
in the ADHM construction constitute a basis of the section space
of $M$. So the matrix $U$ consists of orthonormal basis vectors of
the section space of the instanton bundle $L$, and $UU^\dag$ is
the projection operator corresponding to $L$. As is familiar to
us, the gauge potential (\ref{A by U}) is natural as the
Levi-Civita connection on $L$. The above statements briefly
explain how the instanton bundle can be related to the familiar
ADHM objects.

To formulate the Drinfeld-Manin construction in the ADHM language,
we first concentrate on the $U(2k)$ $k$-instanton case. Now
\begin{equation}
h=\left(\begin{array}{cc}
b & a \\
\end{array}\right)
\end{equation}
is a $4k\times 4k$ square matrix, and
\begin{equation}
\Delta=h X,
\end{equation}
where
\begin{equation}
X\equiv\left(\begin{array}{c}
\bar{x}\otimes 1_k \\
1_{2k} \\
\end{array}\right).
\end{equation}
Thus we have
\begin{equation}
\Delta^\dag\Delta=X^\dag h^\dag h X=X^\dag\left(\begin{array}{cc}
1_{2k} & \underline{a} \\
\underline{a}^\dag & a^\dag a \\
\end{array}\right)X\equiv X^\dag Q X,
\end{equation}
where
\begin{equation}\label{UL a}
\underline{a}\equiv\left(\begin{array}{cc}
B_{2}^{\dagger } & -B_{1} \\
B_{1}^{\dagger } & B_{2}
\end{array}\right)
\end{equation}
is the lower blocks of $a$.

In fact, the column vectors of $X$ constitute a basis of the
section space of $M$ (before we perform the complex linear
transformation mentioned above) and $Q$ is the corresponding
metric. From the ADHM point of view now, to make
$\Delta^\dag\Delta$ of the factorized form (\ref{factorize}), it
is easy to see that $Q$ must satisfy the following factorization
condition:
\begin{equation}\label{on a}
a^\dag a=\left(\begin{array}{cc}
R & 0 \\
0 & R \\
\end{array}\right),
\end{equation}
where $R$ is a $k\times k$ constant Hermitian matrix. Using the
auxiliary symmetry transformation (\ref{aux}), we can make $R$
diagonalized:
\begin{equation}\label{diag}
R=\mathrm{diag}(r_1,r_2,\cdots,r_k),\quad r_1\leq
r_2\leq\cdots\leq r_k.
\end{equation}
On the one hand, we can assume the above form of $R$ to fix the
auxiliary symmetry, which is nonphysical; on the other hand, even
assuming this cannot completely fix the auxiliary symmetry: for
generic $R$ this residual symmetry is $U(1)^{\times k}/U(1)$, and
if some of the $r_i$ are equal, this residual symmetry is even
larger. Further, for generic $R$ this residual symmetry can be
completely fixed by requiring $(k-1)$ of the off-diagonal elements
of $B_1$ or $B_2$, say $(B_1)_{ik}\ (i=1,2,\cdots,k-1)$, to be
real; special cases of coincident $r_i$ can be carefully treated
as well.

To sum up, we can choose $\underline{a}$ and $R$ of the form
(\ref{diag}) as the collective coordinates of the $U(2k)$
$k$-instanton, while removing some of the degrees of freedom in
$\underline{a}$. Obviously, the number of independent real
parameters is $4k^2+k-(k-1)=4k^2+1$, which coincides with the
parameter counting in last section. Noting that
\begin{equation}
a^\dag a=\underline{a}^\dag\underline{a}+K^\dag K,
\end{equation}
where
\begin{equation}
K\equiv\left(\begin{array}{cc}
I^\dag & J \\
\end{array}\right)
\end{equation}
is the upper blocks of $a$, $\underline{a}$ and $R$ must satisfy
the condition that
\begin{equation}\label{S}
S\equiv 1_2\otimes R-\underline{a}^\dag\underline{a}
\end{equation}
is a positive semidefinite matrix. This condition introduces a
boundary to the span of the parameters in $\underline{a}$ and $R$.
Thus we have obtained parametrization of the $U(2k)$ $k$-instanton
moduli space, though the complicated boundary makes it a little
imperfect, which is an inevitable consequence of the highly
nontrivial topology of the instanton moduli space. This
parametrization (also for the following $N>2k$ case) is, in fact,
rediscovered by Dorey \textit{et al.} in 1999 \cite{DHKMV}, but
they did not point out the relation between their work and \cite{D
M}.

Now the matrix $K$ can be expressed in terms of $\underline{a}$
and $R$ due to
\begin{equation}\label{K}
K^\dag K=S.
\end{equation}
Because in the present case $K$ is a square matrix, one may
naturally take $K=K^\dag=S^{1/2}$. This expression of $K$ seems
simple and explicit, but it includes three steps: diagonalizing,
extracting the square root, and undoing the diagonalization. In
fact, to diagonalize $S$ needs to solve a equation of degree $k$,
which we must avoid if we have better choices. Fortunately, a
better choice does exist. We may have in remembrance the
simplification of quadratic forms via congruent transformations in
basic linear algebra:
\begin{equation}
B^\mathrm{T}E B=A,
\end{equation}
where $E$ is the canonical form of $A$. If $A$ is nonsingular, $E$
will be the identity matrix. Otherwise $E$ will have the form
$\mathrm{diag}(1,\cdots,1,0,\cdots,0)$, which can be considered,
in a different point of view, as $E$ always being the identity
while allowing $B$ to be singular:
\begin{equation}
B^\mathrm{T}B=A.
\end{equation}
The transformation matrix $B$ can be easily obtained by completing
squares or by simultaneous row and column transformations, without
solving any nonlinear equations. Now $S$ here is a Hermitian form,
not a quadratic one, but the method is similar.

Next we can consider the $N\ne 2k$ cases. These are very simple.
If $N>2k$, it is easy to find, as has been shown in many
literatures, a natural embedding of the above $U(2k)$ solution $K$
in the $U(N)$ solution $K'$:
\begin{equation}
K'=\left(\begin{array}{c}
0_{(N-2k)\times 2k} \\
K \\
\end{array}\right).
\end{equation}
This gives the $4k^2+1$ ``purely'' physical degrees of freedom of
the $U(N)$ $k$-instanton. To get all the ``ADHM'' degrees of
freedom, i.e., including the global gauge rotations, we only need
to perform the following transformations:
\begin{equation}
K'\rightarrow\mathcal{U}K',\quad\mathcal{U}\in\frac{U(N)}{U(1)\times
U(N-2k)},
\end{equation}
which add $N^2-(N-2k)^2-1$ more parameters to the ``purely''
physical degrees of freedom and make the total number of real
parameters $4Nk$.

If $N<2k$, we can simply restrict the rank of $S$ not greater than
$N$. Then from Eq.(\ref{K}) it is easy to see that $K$ can take
the following form:
\begin{equation}
K=\left(\begin{array}{c}
K' \\
0_{(2k-N)\times 2k} \\
\end{array}\right),
\end{equation}
where the $N\times 2k$ matrix $K'$ is the ADHM matrix for the
$U(N)$ $k$-instanton. Linear algebra tells us that for an $l\times
l$ Hermitian matrix $H$ the condition $\mathrm{rank}(H)\leq l-r$
is equivalent to $r^2$ real conditions on the elements of $H$. So
the number of ``purely'' physical parameters is
$4k^2+1-(2k-N)^2=4Nk-N^2+1$, which again coincides with the
parameter counting in last section. The global gauge rotations are
introduced as
\begin{equation}
K'\rightarrow\mathcal{U}K',\quad\mathcal{U}\in SU(N),
\end{equation}
which supply the other $N^2-1$ real parameters for all the
``ADHM'' degrees of freedom. So far, everything seems well, but in
fact the $(2k-N)^2$ real conditions become another trouble. The
appendix of this paper will show how to explicitly write down
these conditions on elements of $\underline{a}$ and $R$. There we
will see that for $N<2k-1$ they are too complicated to solve, so
this formulism is not appropriate to give explicit solutions for
this case. However, these systematic conditions can be useful to
offer an indirect way to the instanton collective coordinate
integral, which is left for future works.

Only the $N=2k-1$ case is simple. In this case there is only one
condition:
\begin{equation}\label{det}
\det(S)=0,
\end{equation}
which from Eq.(\ref{S}) can be regarded as a quadratic equation of
one of the $r_i$, say $r_k$. So we can take the same free
parameters as in the $N=2k$ case except $r_k$, and express $r_k$
in terms of the other parameters. The quadratic equation
(\ref{det}) has two roots. A little thought will make it clear
that one of the eigenvalues of $S$ has been negative when we take
the smaller root. Thus we can only take the greater one as $r_k$,
which accomplishes parametrization of the $U(2k-1)$ $k$-instanton
moduli space.

\section{Noncommutative Drinfeld-Manin instanton}\label{NCDM}

How to establish the Drinfeld-Manin construction in the
noncommutative case is an interesting problem. Appealing to the
well-developed ADHM construction may be much easier than
considering noncommutative instanton bundles. The commutative ADHM
construction can be regarded as a special case ($\zeta=0$) of the
noncommutative ADHM construction. So we can anticipate that it is
straightforward to generalize the ADHM formulism of the
Drinfeld-Manin construction to the noncommutative case.

In fact, like Eq.(\ref{on a}), the factorization condition
(\ref{factorize}) in the noncommutative case gives the following
condition on $a$:
\begin{equation}
a^\dag a=\left(\begin{array}{cc}
R+\zeta & 0 \\
0 & R \\
\end{array}\right)=\mathrm{diag}(r_1+\zeta,\cdots,r_k+\zeta,r_1,\cdots,r_k).
\end{equation}
So we can similarly choose $\underline{a}$ and $r_i$
($i=1,2,\cdots,k$) as the collective coordinates of the
noncommutative $U(2k)$ $k$-instanton (while removing some of the
degrees of freedom in $\underline{a}$ as in the commutative case).
Now Eq.(\ref{S}) becomes
\begin{equation}\label{NC S}
S\equiv\left(\begin{array}{cc}
R+\zeta & 0 \\
0 & R \\
\end{array}\right)-\underline{a}^\dag\underline{a},
\end{equation}
and the following things are the same as in the commutative case.

To be more clear, our solution of the noncommutative ADHM $U(2k)$
$k$-instanton is
\begin{equation}
a=\left(\begin{array}{c}
S^{1/2} \\
\underline{a} \\
\end{array}\right),
\end{equation}
where $S$ is defined in Eq.(\ref{NC S}) and $\underline{a}$
defined in Eq.(\ref{UL a}). And we must keep in mind that the
square root here is understood in the sense of the simplification
of Hermitian forms, as explained in last section. It is easy to
check that this solution does satisfy the corresponding ADHM
constraints, and it has the correct number of free parameters, as
we have mentioned above.

The techniques to deal with the $N\ne 2k$ cases in the
noncommutative case and that in the commutative case are
completely the same. In fact the global gauge rotations in gauge
field theory are unaffected by the space-time noncommutativity.
Again the $N=2k-1$ case is simple enough to be solved. So we also
obtain parametrization of the noncommutative $U(N)$ ($N\geq 2k-1$)
$k$-instanton moduli space.

To end this paper, let us focus on the two-instanton case. For
$k=2$, we essentially obtain explicit general solutions of the
(commutative or noncommutative) ADHM constraints for $U(N)$
($N\geq 3$) gauge groups. Counting the $U(2)$ two-instanton
solution already known \cite{C W S, Tian4}, we have general
solutions of all the commutative $U(N)$ two-instantons. However,
the general solution of the noncommutative $U(2)$ two-instanton,
which may be of much interest, is yet to be found.

\section*{Acknowledgments}

I would like to thank Prof.~Chuan-Jie Zhu and Prof.~Xing-Chang
Song for helpful discussions. This work is partly supported by
NSFC under Grants No. 10347148.

\appendix

\section{Conditions for a Hermitian matrix of restricted rank}

Consider an $l\times l$ Hermitian matrix $H$. We introduce the
following decomposition of $H$:
\begin{equation}\label{decomp}
H=\left(\begin{array}{cc}
F_{r\times r} & C \\
C^\dag & \underline{H}_{(l-r)\times(l-r)} \\
\end{array}\right),
\end{equation}
and define an $(l-r+1)\times(l-r+1)$ matrix
\begin{equation}
H'_{ij}=\left(\begin{array}{cc}
F_{ij} & C_i \\
C_j^\dag & \underline{H} \\
\end{array}\right),
\end{equation}
where $C_i$ is the $i$th row of $C$. Assuming
$\det(\underline{H})\ne 0$, then the following two propositions
are equivalent:
\begin{itemize}
  \item $\mathrm{rank}(H)=l-r$;
  \item $\det(H'_{ij})=0$ for all $i,j=1,2,\cdots,r$.
\end{itemize}
It is apparent that the latter can be deduced from the former. Now
we explain how the former can be deduced from the latter.

First, for a fixed $j$, the $(l-r)\times(l-r+1)$ matrix
\begin{equation}
\underline{H}'\equiv\left(\begin{array}{cc} C_j^\dag &
\underline{H}
\end{array}\right)
\end{equation}
is obviously of rank $l-r$. Then $\det(H'_{ij})=0$ means that the
rank will not increase when we append a row
$C'_i\equiv\left(\begin{array}{cc} F_{ij} & C_i
\end{array}\right)$ to $\underline{H}'$, so $C'_i$ is a linear
combination of the row vectors of $\underline{H}'$. This is the
case for all $i$, so we can conclude that the following matrix
\begin{equation}
H_j\equiv\left(\begin{array}{cc}
F_j & C \\
C_j^\dag & \underline{H} \\
\end{array}\right)
\end{equation}
is of rank $l-r$, where $F_j$ is the $j$th column of $F$.

Next, the $l\times(l-r)$ matrix
\begin{equation}
H'\equiv\left(\begin{array}{c}
C \\
\underline{H} \\
\end{array}\right)
\end{equation}
is again of rank $l-r$. Thus $\mathrm{rank}(H_j)=l-r$ means that
the rank will not increase when we append a column
\begin{equation}
\hat{C}_j\equiv\left(\begin{array}{c}
F_j \\
C_j^\dag \\
\end{array}\right)
\end{equation}
to $H'$, so $\hat{C}_j$ is a linear combination of the column
vectors of $H'$. Again this is the case for all $j$, so we attain
the desired result $\mathrm{rank}(H)=l-r$.

Because $H$ is Hermitian, $\det(H'_{ij})=0$ are in fact $r^2$ real
conditions. The combination of $\det(\underline{H})\ne 0$ and
these conditions is a sufficient condition for
$\mathrm{rank}(H)\leq l-r$. Of course, it is not necessary. If
$\det(\underline{H})=0$ for the decomposition (\ref{decomp}), we
must take another $(l-r)\times(l-r)$ submatrix of $H$ as
$\underline{H}$ and obtain another $r^2$ real conditions. If $H$
has no nonsingular $(l-r)\times(l-r)$ submatrix, the rank of $H$
is less than $l-r$. Altogether, the requirement
$\mathrm{rank}(H)\leq l-r$ is achieved.


\begin{thebibliography}{99}

\bibitem{Coleman} S. Coleman, {\it Uses of Instantons}, Cambridge University Press (1985).

\bibitem{Rajaraman} R. Rajaraman, {\it Solitons and Instantons}, North-Holland (1982).

\bibitem{S S} T. Sch\"afer and E.V. Shuryak, {\it Instantons in QCD}, Rev. Mod. Phys. {\bf 70} (1998) 323.

\bibitem{S V} M.A. Shifman and A.I. Vainshtein, {\it Instantons versus supersymmetry}, in {\it ITEP Lectures on Particle Physics and Field Theory} (World Scientific, Singapore, 1999).

\bibitem{F U} D.S. Freed, K.K. Uhlenbeck, {\it Instantons and four-manifolds}, Springer-Verlag, New York (1984).

\bibitem{D K} S.K. Donaldson and P.B. Kronheimer, {\it The Geometry of Four-Manifolds}, Clarendon Press, Oxford (1990).

\bibitem{BPST} A.A. Belavin, A.M. Polyakov, A.S. Shvarts and Y.S. Tyupkin, Phys. Lett. B {\bf
59} (1975) 85.

\bibitem{ADHM} M.F. Atiyah, N.J. Hitchin, V.G. Drinfeld and Y.I. Manin, {\it Construction of Instantons}, Phys. Lett. A {\bf 65}
(1978) 185.

\bibitem{C W S} N.H. Christ, E.J. Weinberg and N.K. Stanton, {\it General self-dual Yang-Mills solutions}, Phys. Rev. D {\bf 18} (1978) 2013-2025.

\bibitem{K Sh} V.E. Korepin and S.L. Shatashvili, {\it Rational parametrization of the three-instanton solutions of the Yang-Mills equations}, Sov. Phys. Dokl. {\bf 28}
(1983) 1018-1019.

\bibitem{Tian4} Y. Tian, {\it Comments on the U(2) ADHM two-instanton}, Phys. Lett. B {\bf 566} (2003) 183-185.

\bibitem{Tian3} Y. Tian, C.-J. Zhu and X.-C. Song, {\it Topological Charge of Noncommutative ADHM Instanton}, Mod. Phys. Lett. A {\bf 18} (2003) 1691-1703 [hep-th/0211225].

\bibitem{DHKMV} N. Dorey, T.J. Hollowood, V.V. Khoze, M.P. Mattis and S. Vandoren,
{\it Multi-Instanton Calculus and the AdS/CFT Correspondence in
$N=4$ Superconformal Field Theory}, Nucl. Phys. B \textbf{552}
(1999) 88-168 [hep-th/9901128]; N. Dorey, T.J. Hollowood, V.V.
Khoze, M.P. Mattis, \textit{The Calculus of Many Instantons},
Phys. Rept. \textbf{371} (2002) 231-459 [hep-th/0206063].

\bibitem{K S} A. Konechny and A. Schwarz, {\it Introduction to M(atrix) Theory and Noncommutative Geometry, Part
II}, Phys. Rept. {\bf 360} (2002) 353-465 [hep-th/0107251].

\bibitem{D N} M.R. Douglas and N.A. Nekrasov, {\it Noncommutative Field Theory}, Rev. Mod. Phys. {\bf 73} (2002) 977
[hep-th/0106048].

\bibitem{Szabo} R.J. Szabo, {\it Quantum Field Theory on
Noncommutative Spaces}, Phys. Rept. {\bf 378} (2003) 207-299
[hep-th/0109162].

\bibitem{S W} N. Seiberg and E. Witten, {\it String Theory and Noncommutative Geometry}, J. High Energy Phys. {\bf 9909} (1999)
032 [hep-th/9908142].

\bibitem{Nakajima} H. Nakajima, {\it Heisenberg Algebra and Hilbert
Schemes of Points on Projective Surfaces}, alg-geom/9507012 and
{\it Lectures on Hilbert Schemes of Points on Surfaces}.

\bibitem{N S} N. Nekrasov and A. Schwarz, {\it Instantons on
Noncommutative $R^4$, and (2,0) Superconformal Six Dimensional
Theory}, Commun. Math. Phys. {\bf 198} (1998) 689
[hep-th/9802068].

\bibitem{Tian2} Y.~Tian and C.-J.~Zhu, {\it Remarks on the noncommutative ADHM construction}, Phys. Rev. D {\bf 67} (2003) 045016 [hep-th/0210163].

\bibitem{Witten} E. Witten, {\it Small Instantons in String Theory}, Nucl. Phys. B {\bf 460} (1996) 541.

\bibitem{Douglas} M.R. Douglas, {\it Branes within Branes} [hep-th/9512077].

\bibitem{Moore} M.R. Douglas and G. Moore, {\it D-branes, Quivers and ALE Instantons} [hep-th/9603167].

\bibitem{LTY} K. Lee, D. Tong and S. Yi, Phys. Rev. D {\bf 63} (2001) 065017 [hep-th/0008092].

\bibitem{D M} V.G. Drinfeld and Y.I. Manin, Commun. Math. Phys. {\bf 63} (1978) 177.

\bibitem{Nekrasov} N. Nekrasov, {\it Trieste Lectures on Solitons in
Noncommutative Gauge Theories} [hep-th/0011095].

\bibitem{Connes} A. Connes, {\it Noncommutative Geometry} (Academic Press, 1994).

\bibitem{Madore} J. Madore, {\it An Introduction to Noncommutative Geometry and its Physical Applications} (Cambridge University Press, 1999).

\bibitem{M R S} S. Minwalla, M. Van Raamsdonk and N. Seiberg, {\it Noncommutative perturbative
dynamics}, JHEP {\bf 0002} (2000) 020 [hep-th/9912072].

\bibitem{Furuuchi} K. Furuuchi, {\it Instantons on noncommutative $R^4$ and projection operators}, Prog. Theor. Phys. {\bf 103} (2000) 1043 [hep-th/9912047].

\end{thebibliography}
\end{document}